\documentclass[preprint,showpacs,preprintnumbers,amsmath,amssymb,graphicx]{revtex4}
\usepackage{amsmath,amsfonts,amssymb}
\usepackage{epsfig}
\def\bit{\begin{itemize}}
\def\eit{\end{itemize}}
\def\ben{\begin{enumerate}}
\def\een{\end{enumerate}}
\def\bed{\begin{description}}
\def\eed{\end{description}}

\def\lsim{\raise0.3ex\hbox{$<$\kern-0.75em\raise-1.1ex\hbox{$\sim$}}}
\def\gsim{\raise0.3ex\hbox{$>$\kern-0.75em\raise-1.1ex\hbox{$\sim$}}}

\let\jnfont=\rm
\def\NPB#1,{{\jnfont Nucl.\ Phys.\ B }{\bf #1},}
\def\PLB#1,{{\jnfont Phys.\ Lett.\ B }{\bf #1},}
\def\EPJC#1,{{\jnfont Eur.\ Phys.\ Jour.\ C }{\bf #1},}
\def\PRD#1,{{\jnfont Phys.\ Rev.\ D }{\bf #1},}
\def\PRL#1,{{\jnfont Phys.\ Rev.\ Lett.\ }{\bf #1},}
\def\MPLA#1,{{\jnfont Mod.\ Phys.\ Lett.\ A }{\bf #1},}
\def\JPG#1,{{\jnfont J.\ Phys.\ G}{\bf #1},}
\def\CTP#1,{{\jnfont Commun.\ Theor.\ Phys.\ }{\bf #1},}
\def\JHEP#1,{{\jnfont JHEP \ }{\bf #1},}
\def\NPPS#1,{{\jnfont Nucl.\ Phys.\ Proc.\ Suppl.\ }{\bf #1},}

\def\beq{\begin{equation}}
\def\eeq{\end{equation}}
\def\bea{\begin{eqnarray}}
\def\eea{\end{eqnarray}}
\newcommand{\ba}{\begin{array}}
\newcommand{\ea}{\end{array}}

\begin{document}

\title{Current experimental constraints on the lightest Higgs boson mass
in the constrained MSSM}

\author{Junjie Cao$^{1,2}$, Zhaoxia Heng$^1$, Dongwei Li$^1$, Jin Min Yang$^3$ }

\affiliation{
$^1$ Department of Physics, Henan Normal University, Xinxiang 453007, China\\
$^2$ Center for High Energy Physics, Peking University,
       Beijing 100871, China \\
$^3$  State Key Laboratory of Theoretical Physics, \\
     Institute of Theoretical Physics, Academia Sinica,
             Beijing 100190, China
\vspace*{.8cm} }

\begin{abstract}
We examine the parameter space of the constrained MSSM by
considering various experimental constraints.
For the dark matter sector, we require the neutralino dark matter
to account for the relic density measured by the WMAP and satisfy
the XENON limits on its scattering rate with the nucleon.
For the collider constraints, we consider all relevant direct and
indirect limits from LEP, Tevatron and LHC as well as the muon
anomalous magnetic moment. Especially, for the limits from
 $B_s\to \mu^+\mu^-$, we either directly consider its branching ratio
with the latest LHC data or alternatively consider the double ratio
of the purely leptonic decays defined by
$\frac{Br(B_s\to \mu^+\mu^-)/Br(B_\mu\to \tau\nu_\tau)} {Br(D_s\to
\tau \nu_\tau)/Br(D\to \tau \mu_\tau)}$. We find that under these constraints,
the mass of the lightest Higgs boson ($h$) in both the CMSSM and the NUHM2
is upper bounded by about 124 GeV (126 GeV)  before (after) considering its theoretical uncertainty.
We also find that for these models the di-photon Higgs signal at the LHC is
suppressed relative to the SM prediction, and that the lower bound of
the top-squark mass goes up with $m_h$,  reaching 600  GeV for $m_h=124$ GeV.
\end{abstract}
\maketitle

As a corner stone of the Standard Model (SM), the Higgs boson is now
being exhaustively hunted at the LHC. Very recently both the CMS and
ATLAS collaborations  reported some hints for a relatively light
Higgs boson with its mass at 124 GeV \cite{CMS-PAS-HIG-11-032} and 126
GeV \cite{ATLAS-CONF-2011-163} respectively. Such a light Higgs
boson can be neatly accommodated both in the SM (the electroweak
precision data require a Higgs boson lighter than about 160 GeV) and
in low energy supersymmetric models which predict a rather light
Higgs boson below 135 GeV. Due to the large number of the free
parameters, the minimal supersymmetric standard model (MSSM) can
easily predict a Higgs boson with its mass around 125 GeV
\cite{125Higgs}. The constrained MSSM, however, may not be so easy
to give such a mass due to its rather restrictive parameter space \cite{Const-MSSM}.
In this note we examine the Higgs boson mass in the constrained MSSM
by considering various experimental constraints on its parameter
space. \vspace*{.4cm}

We start our analysis with a description of the constraints we investigate,
which arise from both dark matter experiments and collider experiments. For
the dark matter sector, we require the neutralino dark matter
to account for the relic density measured by the WMAP
($0.1053<\Omega_{CDM}h^2<0.1193$) \cite{wmap} and satisfy the
XENON limits ( 90\%C.L.) on its scattering rate with the nucleon \cite{XENON}.
As shown in \cite{cao-dm}, a large part of the parameter space in
low energy SUSY models can be excluded by such limits.
For the collider constraints, we consider the following as in \cite{cao-dm}:
(1) The LEP search for Higgs bosons; (2) The LEP limits on the masses of
the sparticles such as charginos, sleptons and the third generation
squarks \cite{PDG}, and the limits on the productions of charginos and neutralinos;
(3) The Tevatron limits on charged Higgs bosons in decays of top quarks \cite{Tevatron-charged}.
We do not consider the Tevatron limits on the masses of sparticles because they are
generally weaker than the LHC limits in the constrained MSSM \cite{searchSUSY};
(4) The LHC search for SUSY Higgs bosons via
 $H/A\to \tau\bar\tau$ (95\% C.L.) \cite{CMS-HIG-11-029};
(5) The latest SUSY search results (95\% C.L.) at the LHC \cite{searchSUSY}.
(6) The limits from the electroweak precision observables
including $R_b$ at $2\sigma$ \cite{Cao};
(7) The discrepancy of the SM prediction of muon g-2
from its experimental value, $\delta a_\mu = (25.5\pm 8.2)\times 10^{-10}$ \cite{mug-2}.
We require SUSY to explain the discrepancy at $2\sigma$ level;
(8) The limits from various low energy processes \cite{SuperIso}:
\begin{eqnarray*}
&& 2.0\times10^{-7}<Br(B \to X_s\mu^+\mu^-)<6.8\times10^{-7},\\
&& 2.16\times10^{-4}<Br(B \to X_s\gamma)<4.93\times10^{-4},\\
&& 4.7\times10^{-2}<Br(D_s\to\tau\nu_\tau)<6.1\times10^{-2},
\end{eqnarray*}
\begin{eqnarray*}
&& 3.0\times10^{-4}<Br(D\to\mu\nu_\mu)<4.6\times10^{-4}, \\
&& 0.71\times10^{-4}<Br(B_u\to\tau\nu_\tau)<2.57\times10^{-4};
\end{eqnarray*}
(9) Because $Br(B_s\to\mu^+\mu^-) \propto m_t^4 \mu^2 A_t^2 \tan^6 \beta/(m_A^4 M_{\tilde{t}}^4)$
with $M_{\tilde{t}}$ and $m_A$ denoting the mass scale of top-squark and the CP-odd Higgs
boson mass respectively \cite{Bsmumu}, and thus it is able to serve as a sensitive probe of SUSY with
large $\tan \beta$, and also because recently the experimental upper bound on
$Br(B_s\to\mu^+\mu^-) $ was greatly improved \cite{LHC}, we pay special attention to this quantity.
Especially, noting that $Br(B_s\to\mu^+\mu^-) \propto  f_{B_s}^2 $ with the decay constant $f_{B_s}$
subject to large theoretical uncertainty, we consider the double ratio of purely leptonic decays defined by \cite{R}
\begin{equation}
R\equiv\frac{\eta}{\eta_{SM}},
\end{equation}
with
\begin{equation}
\eta\equiv\frac{Br(B_s\to\mu^+\mu^-)/Br(B_u\to\tau\nu_\tau)}
{Br(D_s\to\tau\nu_\tau)/Br(D\to\mu\nu_\mu)}.
\end{equation}
As pointed out in \cite{R}, the quantity $R$ is quite theoretically clean, and after considering the latest
LHC upper bound on $Br(B_s\to\mu^+\mu^-)$
($1.1\times10^{-8}$ at $95\%$ C.L. \cite{LHC}), $R$ should be less than 2.3 at $95\%$ C.L..
In our analysis we will require either $Br(B_s\to\mu^+\mu^-)\leq 1.26 \times 10^{-8}$ which has
considered the theoretical uncertainties of $Br(B_s\to\mu^+\mu^-)$ (see the manual of the package
SuperIso \cite{SuperIso}), or $R \leq 2.3$ to take into account the constraint from $B_s \to \mu^+ \mu^-$.

\begin{figure}[t]
\includegraphics[width=10cm]{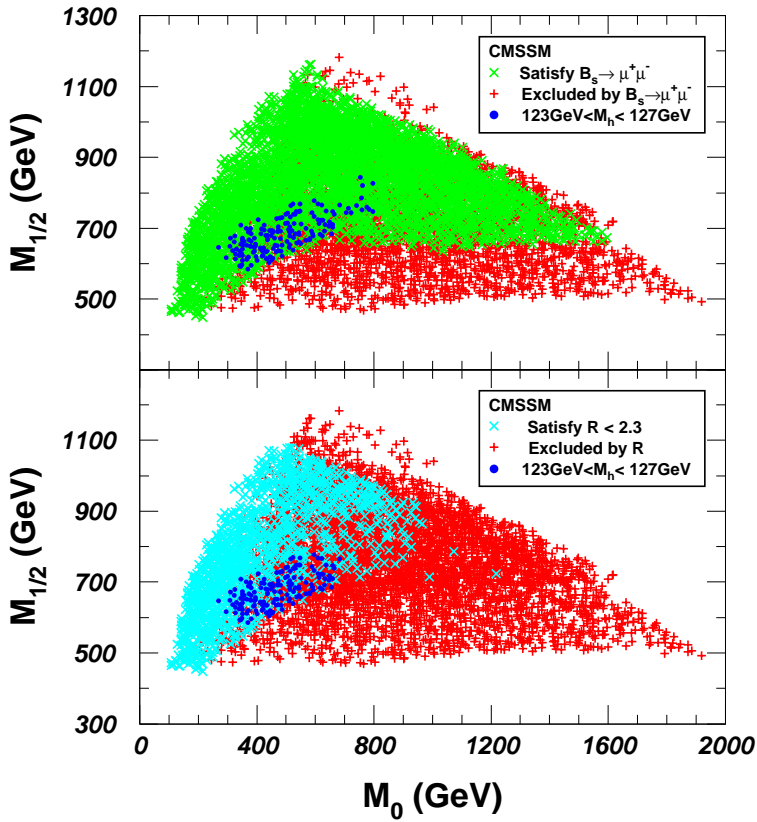}
\vspace{-0.5cm} \caption{The scatter plots of the surviving sample
in the CMSSM, displayed on the plane of $M_0$ versus $M_{1/2}$. In
the upper frame the crosses (red) denote the samples satisfying all
the constraints except $B_s\to\mu^+\mu^-$, and the times (green)
denotes those further satisfying the $Br(B_s\to\mu^+\mu^-)$
constraint. In the lower frame, the crosses (red) are same as those
in the upper frame, while the times (sky-blue) denote the samples
further satisfying the $R$ constraint. In both frames the bullets
(blue) denote the samples with the Higgs boson in the range $123
{\rm~GeV}<m_h <127 {\rm~GeV}$.} \label{fig1}
\end{figure}

\begin{figure}[htbp]
\includegraphics[width=12cm]{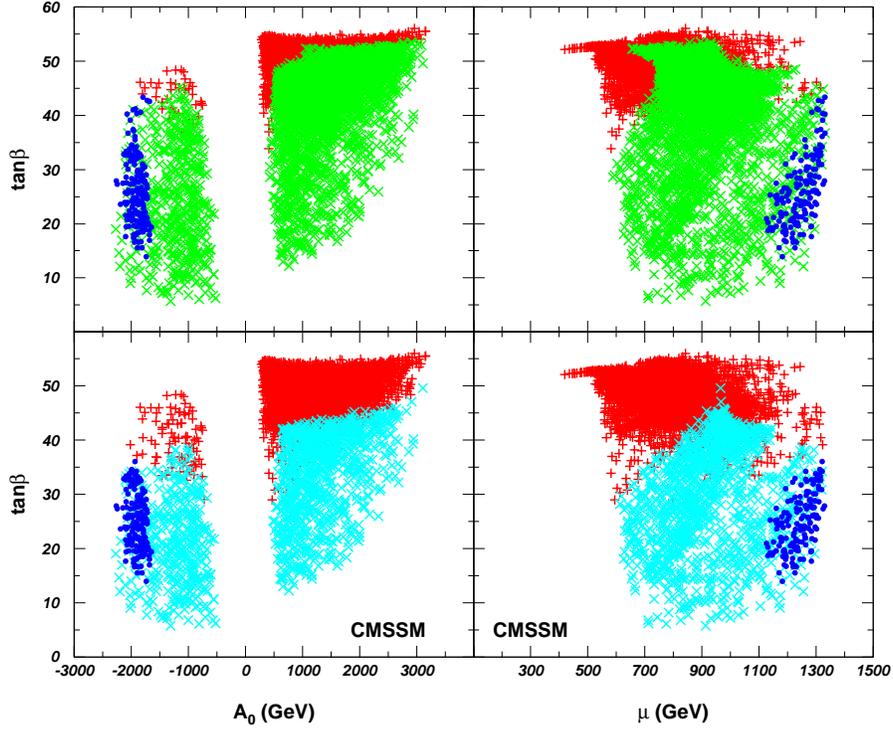}
\vspace{-0.5cm} \caption{Same as Fig.1, but displayed in the $A_0-\tan \beta$ plane
and $\mu-\tan \beta$ plane.} \label{fig2}
\end{figure}

\begin{figure}[htbp]
\includegraphics[width=12cm]{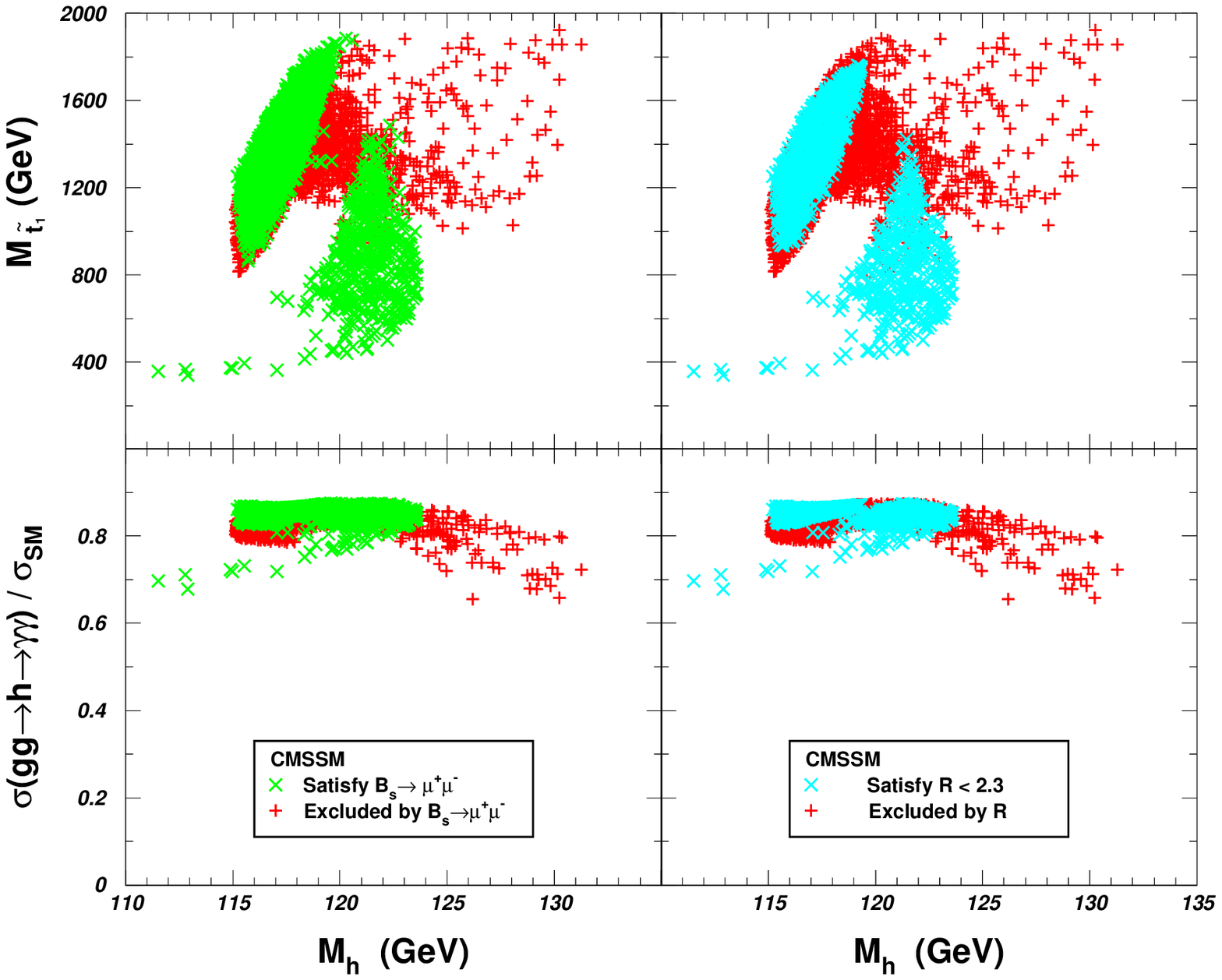}
\vspace{-0.5cm} \caption{Same as Fig.1, but displayed in the planes
of the top-squark mass and the LHC di-photon rate versus the Higgs
boson mass.} \label{fig3}
\end{figure}

In the following, we will first consider the simplest version of the constrained MSSM called CMSSM.
This model is motivated by the paradigm minimal supergravity \cite{mSUGRA}, and
its free parameters consist of $M_0$,~$M_{1/2}$,~$A_0$,~$\tan\beta$ and ${\rm sign}(\mu)$,
where $M_0$ and $M_{1/2}$ are the common scalar mass and gaugino mass respectively, $A_0$ is a
common trilinear soft SUSY breaking parameter, and all of them are defined at the GUT scale.
The parameter $\tan \beta$ represents the ratio of the Higgs field vacuum expectation values,
and $\mu$ is the Higgsino mass. For comparison, we also consider a more
general 2-parameter non-universal Higgs model (NUHM2) inspired by $SU(5)$ grand
unification \cite{nuhm2}.  This model assumes that, beside the input parameters for the CMSSM,
the Higgs soft breaking masses, $M_{H_u}$ and $M_{H_d}$, are all free parameters.
In our analysis, we set ${\rm sign}(\mu)=1$ and vary the CMSSM parameters in the following ranges:
\begin{equation}
 100 {\rm ~GeV}\leq (M_0,  M_{1/2})\leq 2 {\rm ~TeV},
 ~1\leq \tan\beta\leq 60, ~-3 {\rm ~TeV}\leq A_0 \leq 3 {\rm ~TeV}.
\end{equation}
We emphasize that, based on our numerous scan results, only samples within these regions may survive the constraints.
For the NUHM2, motivated by the naturalness of the electroweak symmetry breaking,
we also require
\begin{equation}
|M_{H_u}|, |M_{H_d}| \leq {\rm 1~TeV}.
\end{equation}

In our calculation, we fix $m_t=172.9 {\rm ~GeV}$ \cite{PDG} and $f_{Ts}=0.02$
\cite{lattice} ($f_{Ts}$ denotes the strange quark fraction in the
proton mass). We use the package NMSSMTools \cite{NMSSMTools} in the
MSSM limit (i.e. by choosing very small $\lambda$ and $\kappa$ \cite{Ulrich-report})
to run the soft breaking parameters from the GUT scale
down to the weak scale and implement all the constraints other than
(5), (8) and (9). During the RGE running, the vacuum stability at the weak scale is checked and only the parameter samples that do not spoil the stability 
are kept for further study.
To implement the constraint (5), we note that in the LHC search for SUSY, the $0$-lepton
analyses are in general relatively insensitive to the $\tan \beta$ and $A_0$ parameters in
the CMSSM, and are also insensitive to the amount of Higgs non-universality
in the NUHM \cite{Buchmueller}, so we omit the dependence of the exclusion
bound on the parameters other than $M_0$ and $M_{1/2}$ and take the red
lines in the right panels of Fig.3 and Fig.4 in \cite{searchSUSY} as the
$95\%$ C.L. exclusion limits \cite{Buchmueller,Higgsmassrange}.
For the constraints (8) and (9), we use the package SuperIso \cite{SuperIso}
to study them. After getting the parameter points surviving the
constraints, we calculate the mass of the lightest Higgs ($h$) and
its production rate with the code FeynHiggs \cite{FeynHiggs}.

In our random scan, we have $9.6\times 10^8$ ($8\times 10^8$) samples for the CMSSM (NUHM2),
and obtain 50936 (43194) samples surviving the constraints except (9).  This numbers
is further reduced to 20477 (25978) if the constraint from the latest measurement of
$Br(B_s\to\mu^+\mu^-)$ is added, or alternatively reduced to 6749 (14549) once $R \leq 2.3$ is considered.
This fact reflects that $B_s \to \mu^+ \mu^-$ is able to significantly
limit the constrained MSSM (especially the CMSSM),
and the constraint of $R$ is more stringent than that from $Br (B_s\to\mu^+\mu^-)$.
In the following we will project the surviving
samples in different planes, and in order to show how strong the
constraints are from $B_s\to\mu^+\mu^-$, we will display the samples
without/with the constraint (9).

In Fig.\ref{fig1}-\ref{fig2}, we show the surviving sample of the CMSSM
on the planes of $M_0$ versus $M_{1/2}$, $A_0$ versus $\tan \beta$ and $\mu$ versus
$\tan \beta$ respectively. These figures indicate that significant
parameter regions are excluded by the process $B_s \to \mu^+\mu^-$,
especially samples with $\tan \beta > 52 $ are completely excluded if
the constraint $Br(B_s \to \mu^+ \mu^-) < 1.26 \times 10^{-8}$ is
considered, and samples with $\tan \beta > 45 $ are strongly disfavored
once we require $R \leq 2.3$. For the samples with $R \leq 2.3$,
we found $Br(B_s\to\mu^+\mu^-)$ is usually less than $ 0.8 \times 10^{-8}$,
which explains why the constraint from $R$ is tighter than that directly
from $Br(B_s\to\mu^+\mu^-)$. In these figures, we also show the surviving
samples predicting $123{\rm ~GeV} < m_h < 127 {\rm ~GeV}$. Taking
into account the theoretical uncertainty in calculating $m_h$ and the
experimental error, this mass range is favored by the latest Higgs search
at the LHC \cite{Higgsmassrange}. In this case, the favored parameter regions
are moderate $M_0$ and $M_{1/2}$,  but large $ |X_t| = |A_t - \mu \cot \beta|$.
Since in our scan we only get 166 (upper panel) and 153 (bottom panel) points
with $123{\rm ~GeV} < m_h < 127 {\rm ~GeV}$, we conclude that the
CMSSM will be tightly constrained once $m_h \sim 125 {\rm GeV}$ is
experimentally confirmed in near future.

In Fig.\ref{fig3} we show the correlations of the lighter top-squark mass and
the LHC di-photon rate with $m_h$ in the CMSSM.
We see that after the inclusion of the constraints from $B_s \to \mu^+ \mu^-$, the
upper bound of $m_h$ is reduced from $132{\rm ~GeV}$ to $124 {\rm ~GeV}$.
We will discuss the underlying reason below.
We can also see that, for $m_h \geq 120 {\rm ~GeV}$,
the lower bound of $m_{\tilde{t}_1}$ increases while the upper bound decreases as $h$
becomes heavier.
This can be well understood by the approximate formula of the radiative
correction to $m_h$ \cite{125Higgs}:
\begin{equation}
\Delta m^2_{h
}  \simeq  \frac{3m^4_t}{2\pi^2v^2} ln\frac{M^2_S}{m^2_t} +
\frac{3m^4_t}{2\pi^2v^2} \left( \frac{X^2_t}{M_S^2} - \frac{X^4_t}{12M^4_S}\right)
\end{equation}
where $v=246 {\rm ~GeV}$ and $M_S = \sqrt{m_{\tilde{t}_1}m_{\tilde{t}_2}}$.
This formula tells us that a heavy top-squark (corresponding
the increase of the lower bound) or a large $X_t$ (corresponding to
the decrease of the upper bound) is necessary to push up the Higgs boson mass.
Fig.\ref{fig3} also shows that for $m_h \simeq 124 {\rm ~GeV}$, the value of $m_{\tilde{t}_1}$
varies from $600 {\rm ~GeV}$ to $1 {\rm ~TeV}$.
In this case, the induced fine tuning problem is not so severe.

\begin{figure}[t]
\includegraphics[width=12cm]{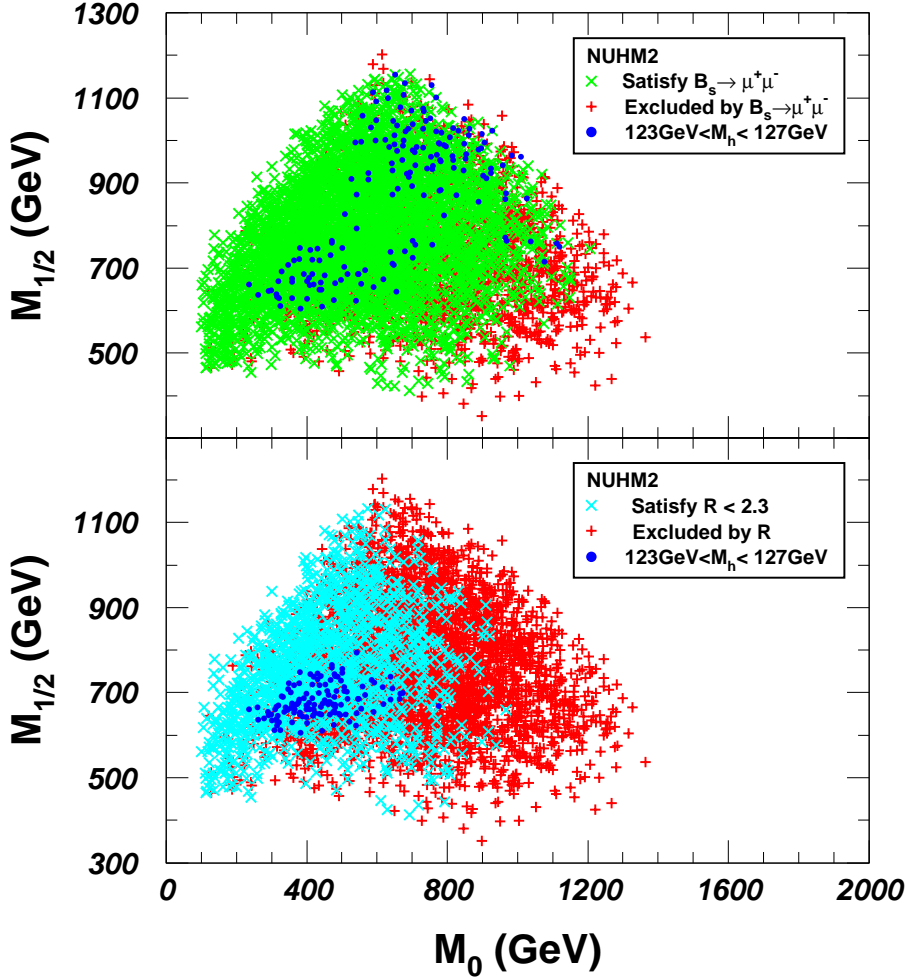}
\vspace{-0.5cm} \caption{Same as Fig.1, but for the NUHM2.}
\label{fig4}
\end{figure}

From Fig.\ref{fig3} we also find that the di-photon Higgs signal at
the LHC is suppressed relative to the SM prediction. We checked that
such a suppression mainly comes from the enhanced $h\bar{b}b$
interaction so that the total width of $h$ is enlarged \cite{di-photon}.
So far the di-photon signal reported by the ATLAS and CMS experiments is
consistent with its SM prediction \cite{CMS-PAS-HIG-11-032,ATLAS-CONF-2011-163},
but due to their large experimental uncertainties,
it is too early to use the di-photon signal rate to exclude the CMSSM.

\begin{figure}[t]
\includegraphics[width=12cm]{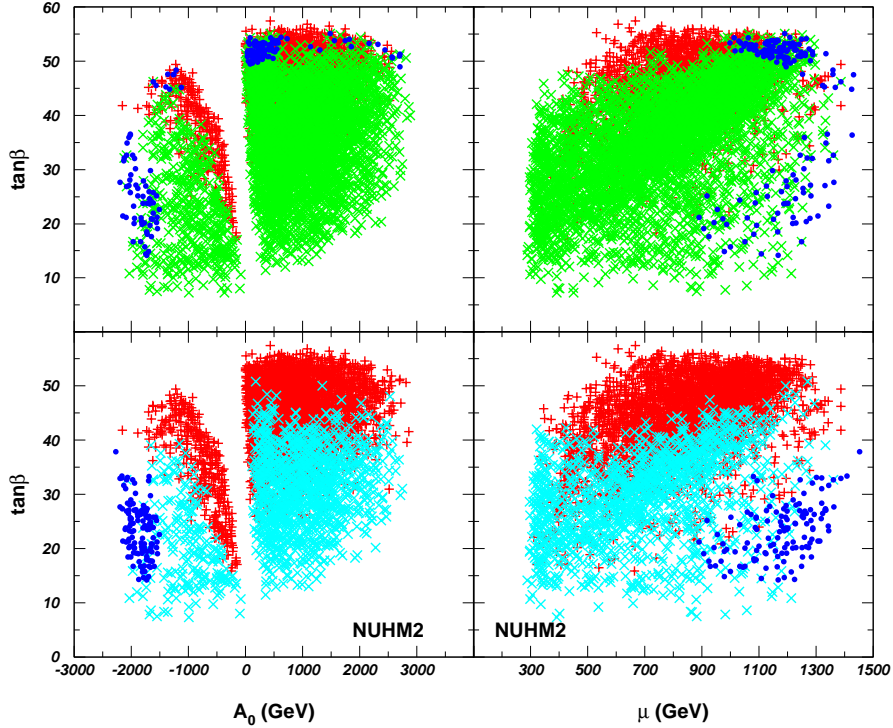}
\vspace{-0.5cm} \caption{Same as Fig.2, but for the NUHM2.}
\label{fig5}
\end{figure}

\begin{figure}[htbp]
\includegraphics[width=12cm]{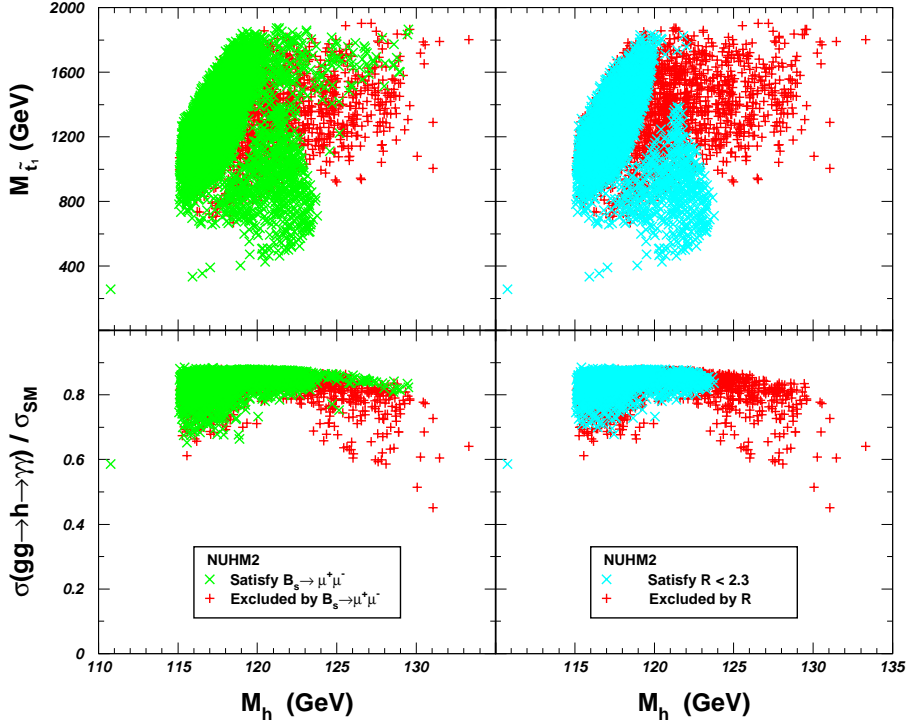}
\vspace{-0.5cm} \caption{Same as Fig.3, but for the NUHM2.}
\label{fig6}
\end{figure}

Since the CMSSM is difficult to predict a Higgs boson with a mass
indicated by the ATLAS and CMS experiments, we consider the NUHM2.
This model is a more general constrained MSSM with two more free
parameters than the CMSSM so that one may tune the value of $m_A$ to
escape the limit from $B_s \to \mu^+ \mu^-$. In Fig.\ref{fig4}-\ref{fig6},
we show our results for the NUHM2. From these figures one can learn
that in the NUHM2 the constraint of $Br(B_s \to \mu^+ \mu^-)$ allows
for relatively heavy SUSY and $m_h$ can be as large as $130{\rm
~GeV}$. But the constraint of $R$ is still stringent, which requires
$m_h<124 {\rm ~GeV}$. For this case, other features, such as the lighter
stop mass, the di-photon rate and the parameter regions to predict
$123{\rm ~GeV} < m_h < 127 {\rm ~GeV}$, are quite similar to those of the CMSSM.

About the above results, we have more explanations. First, in calculating
the scattering rate of the neutralino dark matter with the nucleon, we choose a
relatively small $f_{Ts}$, $f_{Ts} = 0.02$, which is given by the recent
lattice computation \cite{lattice}. In this case, the contribution of the
strange quark in the nucleon to the scattering rate is less important. 
Since a larger $f_{Ts}$ usually enhances the scattering rate \cite{cao-dm1}, 
our choice of $f_{Ts}$ actually leads to a conservative constraint from 
the XENON100 experiment. Second, we note so far the
the top quark mass has sizable experimental uncertainty. We checked that the change
of $m_t$ from $172.9 {\rm GeV}$ to $173.9 {\rm GeV}$ will increase $m_h$ by
less than $0.8 {\rm GeV}$. We also examined the theoretical uncertainty of $m_h$
by FeynHiggs, which may arise from the variation of the
renormalization scale from $m_t/2$ to $2 m_t$, the use of $m_t^{pole}$
instead of $m_t^{run}$ in the two-loop corrections and the exclusion of
higher-order resummation effects in $m_b$ \cite{FeynHiggs}. We found the
uncertainty $\delta m_h$ is usually less than $2.5 {\rm GeV}$, and for
the sky-blue samples shown in bottom panel
of Fig.\ref{fig1} (Fig.\ref{fig4}), only about $110$ (90) points of them 
predict $m_h + \delta m_h $ exceeding $125 {\rm GeV}$, but no point 
exceeds $126 {\rm GeV}$. This again
reflects the difficultly of the CMSSM (NUHM2) in  predicting $m_h \simeq 125 {\rm GeV}$.
Third, we note that at the LHCb with $3 ~fb^{-1}$ integrated luminosity, the rare
decay $B_s \to \mu^+ \mu^-$ can be discovered for a rate down to $7 \times
10^{-9}$ \cite{R}. This will provide a good opportunity in near future
to further test the constrained MSSM. Last, we provide an intuitive
understanding about why the Higgs boson mass $m_h$ is so severely constrained
by the process $B_s\to\mu^+\mu^-$ through emphasizing two facts. One is in the constrained
MSSM the sfermions have a common boundary mass, so the slepton masses are
correlated with the squark masses. Since we require the SUSY effects to explain
the muon $g-2$ at $2\sigma$ level, the sleptons (and thus the squarks) can not
be too heavy given $\tan \beta < 60 $ as required by perturbativity. This means
that heavy squarks must be accompanied by a large $\tan \beta$ in order
to explain the muon $g-2$. The other fact is since $Br(B_s\to\mu^+\mu^-)$ is very
sensitive to $\tan \beta$ (proportional to $\tan^6 \beta  A_t^2$), the constraint
from $B_s\to\mu^+\mu^-$ can restrict tightly the value of $\tan \beta$ and
consequently further restricts the squark masses. Given the limited top-squark
masses, the only way to enhance $m_h$ is through a large $A_t$, which, however,
enhances $Br(B_s\to\mu^+\mu^-)$ and thus get constrained.
 \vspace*{0.5cm}

In summary, under the current experimental constraints at $2\sigma$ level (except
the constraints from the XENON experiment which are at $90\%$ C.L.),
we performed a random scan in the parameter space of the constrained MSSM
and obtained the following observation: (i) the mass of the lightest Higgs boson
($h$) in the CMSSM and NUHM2 is upper bounded by about 124 GeV (126 GeV) before (after)
considering its theoretical uncertainty; (ii) the di-photon Higgs signal at the LHC
is suppressed relative to the SM prediction; (iii) the lower bound of the top-squark
mass goes up with $m_h$ and for $m_h=124$ GeV the
top-squark must be heavier than 600 GeV. Therefore, if the ATLAS
(CMS) Higgs hint around $125 \rm{GeV}$ is the true story, these
models will be tightly constrained. \vspace*{0.5cm}

{\em Note added:} While we were preparing this manuscript, we found
some similar works appeared in the arXiv \cite{Higgsmassrange,CMSSM,CMSSM-NUHM}.
Let us clarify the main difference of our work from those works.
In \cite{Higgsmassrange}, the authors
build a likelihood function by incorporating relevant experimental data
and aim at finding parameter regions favored by
experiments in the framework of the CMSSM; while
in \cite{CMSSM}, the authors investigate how large $m_h$
can reach in the CMSSM without considering seriously
the constraint from the muon $g-2$. Compared with \cite{CMSSM-NUHM}
where the authors investigate $m_h$ in the same models as in our work,
we considered more constraints.

Next, we emphasize again that, although a small $Br(B_s \to \mu^+ \mu^-)$
can be easily accommodated in heavy SUSY, the latest experimental results
on $Br(B_s \to \mu^+ \mu^-)$ can severely constrain the CMSSM parameter
space. This is because we require the SUSY effects to explain the muon
$g-2$, which favors SUSY at moderate scale and large $\tan \beta$. Also we
want to stress again that our observation of the upper bound of 124 GeV on $m_h$
is obtained by scanning about $10^9$ random samples under various experimental
constraints listed at the beginning of this paper.

Finally, during the revision of this manuscript, the updated information on
$B_s \to \mu^+ \mu^-$ at the CMS  appeared as  $Br(B_s \to \mu^+ \mu^-)
\leq 0.77 \times 10^{-8}$ \cite{Bsmumu-CMS}. We checked that with such a 
new limit the $Br(B_s \to \mu^+ \mu^-)$ constraint will be comparable with 
the $R$ constraint presented in our results.
\vspace*{0.5cm}

This work was supported in part by the National Natural
Science Foundation of China (NNSFC) under grant Nos. 10821504,
11135003, 10775039, 11075045, by Specialized Research Fund for
the Doctoral Program of Higher Education with grant No. 20104104110001,
and  by the Project of Knowledge Innovation Program (PKIP) of
Chinese Academy of Sciences under grant
No. KJCX2.YW.W10.

\end{document}